PAPER

# Lessons learned to boost a bioinformatics knowledge base reusability, the Bgee experience

Tarcisio Mendes de Farias[1,2,\*], Julien Wollbrett[1,2], Marc Robinson-Rechavi[1,2] and Frederic Bastian[1,2]

[1]SIB Swiss Institute of Bioinformatics and [2]Department of Ecology and Evolution, University of Lausanne

[\*]tarcisio.mendes@sib.swiss

## Abstract

**Background**, enhancing interoperability of bioinformatics knowledge bases is a high priority requirement to maximize data reusability, and thus increase their utility such as the return on investment for biomedical research. A knowledge base may provide useful information for life scientists and other knowledge bases, but it only acquires exchange value once the knowledge base is (re)used, and without interoperability the utility lies dormant. **Results**, in this article, we discuss several approaches to boost interoperability depending on the interoperable parts. The findings are driven by several real-world scenario examples that were mostly implemented by Bgee, a well-established gene expression database. To better justify the findings are transferable, for each Bgee interoperability experience, we also highlighted similar implementations by major bioinformatics knowledge bases. Moreover, we discuss ten general main lessons learnt. These lessons can be applied in the context of any bioinformatics knowledge base to foster data reusability. **Conclusions**, this work provides pragmatic methods and transferable skills to promote reusability of bioinformatics knowledge bases by focusing on interoperability.

**Key words**: Interoperability; databases; data reusability; ontologies; FAIR principles

## 1 Introduction

Bioinformatics knowledge bases (KB) are often built to serve a specific community of interest. By providing software tools, methods, services and data, these KBs aim to facilitate and to provide the means for their users' work, such as scientific research. Therefore, the notion of reusability is an important aspect to be considered by any bioinformatics KB. Reusability is the capability of a resource to be used multiple times by distinct agents. This statement is a generalization of the data reusability definition in [1].

In applied computing, the relevance of data reusability by computer programs has been highlighted since the late 20th century [1]. More recently, leveraging an efficient discovery and reusability of digital research resources by both machines and humans has been endorsed by the Findable, Accessible, Interoperable and Reusable (FAIR) principles since 2016 [2]. According to Jacobsen et al. [3], findability, accessibility, and interoperability together enable the final goal of trusted, effective and sustained reuse of research resources. This goal is also emphasised by Mons et al. in [4] who show that FAIR principles focus on ensuring that research objects are reusable. Thus, having reusability as the ultimate aim highlights that the FAIR principles are intrinsically related and that, ideally, they should not be dissociated nor seen as independent. The structure and abstraction of the FAIR guiding principles in four main different ones (i.e., F.A.I.R.) has been the source of confusion and misinterpretations by researchers and practitioners trying to "FAIRify" their data and metadata. This can be clearly seen when the original authors needed to explain their initial intent and interpretations of the FAIR principles in later publications [3, 4], or furthermore, when other researchers try to interpret them [5, 6]. This issue is also acknowledged by authors in [7]. In [8], for example, authors state that the operationalization of these principles is yet to be agreed upon within distinct research domains. Nonetheless, despite those misunderstandings, it is undeniable that the FAIR principles significantly contribute to the awareness of different stakeholders (e.g., politicians, research funding agencies, researchers, see [9]) about the relevance of making research outputs reusable by different agents, that is both humans and machines.

In this article, among several similar and complementary definitions of interoperability as reported in [10], we consider the following IEEE standard definition of interoperability: "the ability of two or more systems or elements to exchange information and to use the information that has been exchanged" [11]. Based on this definition, and looking at bioinformatics KBs as systems, we will describe how to improve reusability through interoperability enhancement. It is important to notice that we do not seek in this work to discuss all aspects of the FAIR principles, although we aim at being compliant with the FAIR principles and beyond to achieve an optimal reusability. One of the reasons why we focus on the interoperability aspect is because without it, for instance, it will be hard or even impossible to be findable and accessible. For example, assigning persistent identifiers or descriptive metadata for digital resources, respectively F1 and F2 principles [2], will only be useful if they are interoperable. This is because for a persistent identifier to be resolved, it will need to interoperate with other systems such as the Digital Object Identifier (DOI) resolver. This is also why when talking about DOI, we emphasize it is a persistent *interoperable* identifier, that complies with the ISO 26324:2012 standard [12]. As a result, it is not fortuitously that the word "interoperability" (including its variants and definition) appears more than 50 times in the DOI handbook [13]. Similarly, the produced descriptive metadata will only be useful, for example, to a search engine, if it can be consumed, that is interoperable.

Moreover, achieving interoperability has been well recognised as a complex task by several researchers [14, 15, 16, 17]. Since it is a hard task, it also presents an impediment to the exchange of information among independent bioinformatics KBs. Therefore, to mitigate this issue, in this paper, we focus on pragmatic approaches for interoperability enhancement of bioinformatics KBs. We mainly illustrate these approaches with our experience with our development of Bgee as a more reusable KB.

Bgee is a well-established knowledge base to retrieve and compare gene expression patterns in multiple animal species [18]. It integrates and harmonises multiple data sources that are based on heterogeneous techniques, namely, single-cell RNA-Seq (scRNA-Seq), bulk RNA-Seq, Affymetrix, in situ hybridization, and Expressed Sequence Tags (EST). It is based exclusively on curated healthy wild-type expression data (e.g., no gene knock-out, no treatment, no disease), to provide a comparable reference of normal gene expression. Moreover, the usefulness of the Bgee interoperability practices has been recognized by several researchers such as in [19].

Although we centre on our work on the Bgee use case, the practices, methods and lessons learned discussed here are transferable to other KBs such as those reported in the Nucleic Acids Research Molecular Biology Database Collection [20]. To further demonstrate they are transferable, for each Bgee interoperability experience, we will also highlight, when it is applicable, similar implementations by widely used bioinformatics KBs such as GeneCards (a human gene-centric KB) [21], UniProtKB (a protein-centric KB) [22] and Orthologous MAtrix (OMA, an orthology resource) [23].

Furthermore, the lessons learned could be applied in part to other contexts such as low carbon energy databases (DB) mentioned in [8]. The latter report that energy DBs are significantly heterogeneous and can benefit from common data exchange formats and semantic representations to improve interoberability among these DBs.

## 2 Broad aspects for improving interoperability

Enhancing data/metadata interoperability is the solution for solving data/metadata heterogeneities among the different parts (e.g., systems) between which we seek to exchange information [24, 25]. According to [26], we can categorise these heterogeneities as structural (schema), syntactic (format) and semantic (meaning) heterogeneity. As noted by Alon Y. Halevy in [27], semantic heterogeneity appears whenever there is more than one way to structure a body of data.

To correctly exchange information between different systems, we have to solve the syntactic and semantic heterogeneities, if any, between producer and consumer of this information. By correctly, we mean the information is perceived by the consumer exactly as it is intended by the producer, and the opposite is also true where the information conceived/written by the producer is defined exactly as expected by the consumer. As a result, there is no need for guessing, heuristics or machine learning methods by the consumer/producer to correctly process the exchanged information. To illustrate this, let us consider a semi-structured format for data exchange such as a comma-separated value (CSV) file (i.e., a tabular data format). The 2005 technical standard RFC 4180 [28] formalizes the CSV file format, however there are still multiple syntactical ways to define a CSV file. For instance, the header line, appearing as the first line of the file, is an optional one and there is no explicit manner to identify whether it is present or not, thus care is required by the consumer when importing data. Therefore, the producer and the consumer have to come to an agreement to solve this syntactical heterogeneity to become interoperable, such as is the case for the Google Ads system [29] or the United States National Center for Biotechnology Information (NCBI) LinkOut service [30].

More flexibility implies more heterogeneity. Choosing a more flexible information exchange solution often implies more heterogeneities to solve [27]. Although, a highly constrained, formal and accurate interoperability solution significantly reduce heterogeneity, its adoption may be compromised due to the difficulties of implementation, adaptability and fitness for information being exchanged. For example, describing data from a model (e.g., the Bgee native data model) into another (e.g., the Wikidata [31] data model) can lead to data loss (i.e., partial interoperability) due to semantic heterogeneities. These heterogeneities exist whenever experts with several modelling practices and constraints (e.g., application scope, real-time capabilities, security) can produce different conceptual models to represent the same ensembles of data. Actually, even if an ontology is defined as "a formal, explicit specification of a shared conceptualization" [32] different ontologists can produce different ontologies for a same knowledge domain. For example, more than ten ontologies in the Ontology Lookup Service [33] report a "different" *Gene* concept. To address the latter issue, notably ontology matching and alignment have been recognized as interesting approaches [34, 35]. In general, data, metadata and data schema mappings between the different interoperable elements enable matching/alignment.

Nevertheless, even to define mappings and alignments, a language with a specific vocabulary, syntax and semantics is chosen and applied. For example, we could define alignments with plain English—implying not machine-ready to be read; programming languages—specific-purpose adaptors/translators; the OWL—Web Ontology Language (e.g., owl:sameAs, owl:equivalentClass) [36]; SKOS—Simple Knowledge Organization System vocabulary (e.g., skos:closeMatch) [37]; SWRL—Semantic Web Rule Language (e.g., swrlb:matches, swrl:Imp) [24]; VoIDext—Extended Vocabulary of Interlinked Datasets (e.g., voidext:resourceMapping) [38] and so on. As a result, the heterogeneity problem, and consequently, interoperability issues persist but at another level. When choosing an interoperability solution, that often includes data models, languages, standards for representing the metadata and data, we have to consider different heterogeneity degrees to solve, as for the definition of mappings and alignments. Moreover, depending on the types of heterogeneity, the nature of elements we wish to interoperate, and the interoperability level we want to achieve, one language can be better than another one to declare mappings. For example, if we are dealing with ontology matching, OWL language is not expressive enough to define complex data schema alignments. For instance, if we suppose the existence of three at-tributes/properties genus, species and scientific name, the concatenation of genus (e.g., Homo) and species (e.g., sapiens) implies the species' scientific name (e.g., Homo Sapiens). Then, to define these complex mappings other languages such as SWRL are more appropriate. SWRL and OWL are both logic-based formalisms. Combining them to define complex mappings further allows us to automatically derive new alignments, thanks to inference engines supporting these languages [24]. In some context, to allow different levels of interoperability in terms of precision, it may be crucial to define the nature of the mappings such as reported in the SKOS vocabulary: `skos:closeMatch`, `skos:exactMatch`, `skos:broadMatch`, `skos:narrowMatch` and `skos:relatedMatch`.

Depending on the interoperability aim we want to achieve, a semantic relaxation approach can be applied. Semantic relaxation is the capacity of ignoring semantic and data heterogeneities for the sake of interoperability [38]. For example, when interoperating with different orthology databases (i.e., containing information about corresponding genes in different species), the concepts of genes and proteins can be interchangeably used. This is because some algorithms infer orthologous genes using protein sequences. Hence, we can increase interoperability if some loss of information or of precision are admissible.

## 3 Knowledge base interoperability approaches and practices

We define three different types of interoperability approaches for KBs as follows:

**Definition 1.** *One-side interoperability: one side must strictly comply with the other's procedure to interoperate. There is no or little possible negotiation between interoperable parts. If the other's procedure to interoperate with is based on an independent interoperability procedure, it will be classified as a multi-side interoperability as defined in Def. 3.*

**Definition 2.** *Two-side interoperability: both sides must reconcile with each other, that is establish a common agreement to interoperate.*

**Definition 3.** *Multi-side interoperability: the two or more sides that want to interoperate comply with an independent interoperability procedure. Improvements or changes in this procedure may be upon request, and may or may not be accepted by the third-party organisation or community that maintains the interoperability procedure. This procedure is usually composed of interoperability standards such as Schema.org. Two-side interoperability is considered a multi-side interoperability, if and only if the reconciliation is based on an independent interoperability solution that can be reused by others.*

The classification of an interoperability approach based on those definitions highly depends on the context and timeline. For example, a one- or two-side interoperability can evolve to a multi-side interoperability, if it becomes a standard or part of one that can be reused by others. In addition, a one-side interoperability can become a two-side one, for instance, if both sides want to improve and increase the information exchanged, that is not supported by the existing one-side approach. Another possible scenario is a hybrid approach where more than one interoperability approach is implemented. For example, a multi-side interoperability might not be sufficient or timely to establish the desired degree of interoperability between KBs. In this scenario, a two-side interoperability approach may complement the multi-side one.

Table 1 exemplifies the interoperability approaches defined in Def. 1, Def. 2 and Def. 3. Moreover, this table summarises the use cases involving Bgee as a data producer that are fully described in the next section.

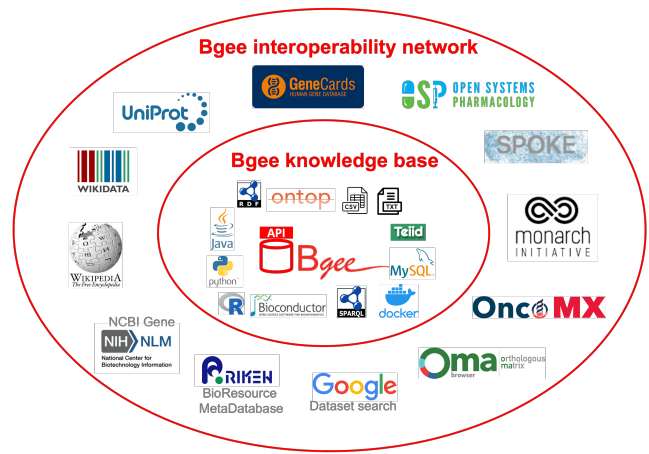

**Figure 1.** A simplified illustration of the boost of the Bgee interoperability network mentioned in this paper. The elements in the inner circle are examples of the ensemble of techniques used to implement the Bgee interoperability with diverse databases and systems, that are illustrated with their logos in the outer circle.

## 4 Enhancing interoperability: the experience of Bgee with other knowledge bases

The Bgee KB integrates and aggregates data from heterogeneous data sources by reconciling them and applying a data warehouse approach, resulting in a large relational database (8TB at time of writing). Curation and quality control are at the core of the Bgee mission. Moreover, being licensed as a public domain database makes Bgee an interesting case study for boosting interoperability since no ownership restrictions exist when reusing its data. Fig. 1 shows a simplified view of the Bgee interoperability network boost and of the technologies involved and Fig. 2 illustrates an overview of the implemented Bgee data interoperability architecture, that are further detailed in the next subsections. Finally, for each Bgee experience we mention similar implementations, if any, by other KBs (namely, UniProtKB, GeneCards, or OMA), and when it is applicable, how they can benefit from our experience too. For instance, if they are not interoperable with a target KB of interest yet.

### 4.1 File-based data exchange

Exchanging data with computer files has been done since the advent of computer file systems. The advantages of this method for exchanging and reusing data among bioinformatics KBs include easy deployment and possible autonomy. For example, the data producer may impose the data format which the data consumers can use without a previous agreement with them. In this example, the consumers have to adapt their tools (e.g., implement a file reader) to be able to interoperate with data producers. Similarly, a data consumer may also impose the data exchange format to be used by data producers. Nevertheless, this interoperation mode often leads to misinterpretations, mainly due to the lack of data interoperability standards and data structure (i.e., unstructured or semi-structure data), and it complicates interoperability because of syntactic and semantic heterogeneities between consumer and producer. Moreover, it does not necessarily provide access to the latest data because of asynchronous and independent exporting and importing data operations of static files.

Currently, Bgee interoperates with the following KBs by file-based data exchange: NCBI Gene database, GeneCards, UniProtKB, RIKEN MetaDataBase, and OncoMX. Moreover, for advanced users, Bgee provides highly structured data through two data dumps using relational and graph data models. In the next subsections, we discuss each Bgee file-based interoperation case and how we mitigate the aforementioned issues such as misinterpretations.

| Target knowledge base (KB) | Interoperability approach | Description |
|---|---|---|
| NCBI Gene [39] | One-side | Bgee must comply with the NCBI LinkOut system exchange file format, that is either a CSV or XML file. |
| UniProtKB [22] | One-side | Bgee must comply with the UniProtKB exchange file format, that is a text file based on its own format. |
| GeneCards [21] | Two-side | Bgee and GeneCards defined from scratch a TSV-like exchange file format that is easy and fast to be consumed by GeneCards and produced by Bgee. |
| OncoMX [40] | Two-side | At first, Bgee and OncoMX defined from scratch a TSV-like exchange file format that was easy and fast to be consumed by OncoMX and produced by Bgee. |
| RIKEN Metadatabase [41] | One-side | RIKEN Metadatabase directly imports the downloadable Bgee RDF dump file into its triple store as a named graph. |
| Monarch Initative [42] | One-side | The Monarch Initative project uses the available Bgee download files as they are. |
| SPOKE [43] | One-side | The Bgee download files are used as they are to build a precision medicine knowledge graph. |
| Open Systems Pharmacology [44] | One-side | The Bgee download files are used as they are to build a KB of gene expression information for drug development. |
| Wikidata [31] | One-side | Bgee developed a bot using Wikidata Python APIs in order to automatically extract, transform and load its data into Wikidata [45]. |
| Wikipedia [46] | One-side | Bgee implemented and integrated a software component in the existing gene infobox module [47]. This allows Wikipedia to dynamically retrieve Bgee data in Wikidata. |
| Google Dataset Search [48] | Multi-side | Bgee provides Schema.org-based metadata embedded in its Web pages. These metadata are automatically retrieved and consumed by other systems that supports Schema.org such as Google Dataset Search tool. |
| OncoMX federation | One-side | OncoMX directly uses the available Bgee MySQL database, called EasyBgee, (i.e., a one-side interoperability) to federate both KBs. To do so, the federated data schema is composed of the OncoMX native relational data schema and a view for the EasyBgee data schema is defined along with mappings. |

**Table 1.** This table summarises the use cases involving Bgee as a data producer along with the implemented interoperability approaches.

*Bgee in the National Center for Biotechnology Information (NCBI) Gene database*

NCBI Gene provides gene-centric information such as sequence, expression, structure, function, citation, and homology data. To be able to interoperate with NCBI Gene through the NCBI LinkOut service [49], we have to strictly comply with NCBI's own specifications to write the expected data-exchange files, either an XML-based file or CSV-based files. As a result, its reader will be able to consume the provided data in a automatic way once deployed at a file-transfer location. The LinkOut system has successfully enabled more than 250 data providers [50], including Bgee, GeneCards and OMA, to link their resources to different NCBI databases such as the NCBI Gene. The data we provide are gene symbols and links to the Bgee gene pages that correspond to a NCBI gene page through the NCBI LinkOut section. Although an extensive documentation is provided for both CSV and XML file definitions [51], the lack of better semantic representations may result in non-compliant files for the NCBI LinkOut file reader by the data producer. Moreover, even though a Document Type Definition (DTD) [52] exists for the LinkOut XML file creation, it does not provide enough control on the XML structure, such as an XML Schema Definition (XSD) [53]. For example, with DTD we are not able to define data types. Relevant XML element data types for the LinkOut XML file definition, such as *<LinkId>* data type, are unknown, and thus, we do not know if a *<LinkId>* can be any character or just integer values greater than zero. Even if the data types or a complete data schema are explicitly defined in the documentation, this schema will not be machine-readable, hence complexifying tasks that could otherwise be automatised, such as the writing of a data exchange file.

To interoperate with the NCBI LinkOut system for publishing the Bgee gene links at NCBI Gene pages, we decided to be compliant with its CSV format. A portion of the generated CSV file is shown in Fig. 3. Although a CSV file provides less structure than an XML one, this decision was mainly dictated by the fact we can easily generate the Bgee tabular data output by simply writing a single Structured Query Language (SQL) query over the Bgee relational database. In addition, in this query, we also project the expected CSV header line by the NCBI LinkOut tool. Nevertheless, the flexibility provided with this file format comes with a high price, that is, the lack of semantics and data structure creating heterogeneities. First, at the syntactic level, the CSV file expected by the LinkOut tool is not fully aligned with most implementations as documented in the standard RFC 4180. This standard formalizes the CSV format and notably states that "each field may or may not be enclosed in double quotes". However, we cannot use quotation marks to enclose fields in the LinkOut CSV file, if we do so, it results into invalid files. Moreover, the LinkOut tool restricts an NCBI Gene entry to have at most three records by the same data provider. Second, at the semantic level, the required third field illustrated in Fig. 3 can have multiple interpretations: either a unique identifier (UID) or a query based on a custom syntax. This field is critical because it enables to intersect the Bgee and NCBI Gene data, in other words, to correctly publish the Bgee links and gene symbols in the NCBI Gene pages. A UID for our use case means the NCBI Gene identifier. In the Bgee relational database, we do not have the NCBI Gene identifiers, hence, we first sought to define a NCBI-like query instead of a UID as a third CSV field by providing common ids between Bgee and NCBI Gene, such as the Ensembl ids. However, this may result in inconsistencies between Bgee and NCBI gene entries, such as an Ensembl id that is not present at NCBI or which does not retrieve the same gene entry. The latter case may occur because the query is a keyword matching any indexed word of a NCBI Gene page. The Bgee team was instructed by the LinkOut service to only consider Ensembl ids present in NCBI Gene, which implies that the LinkOut tool may not work properly if no result is retrieved by a given query (i.e., an Ensembl id keyword). Therefore, in order to avoid ambiguity and to solve this semantic heterogeneity, we decided to use the NCBI Gene ids and to include them in the Bgee database. This was possible thanks to the NCBI Gene id mapping file [54] we imported into our database. It is notable that none of these issues nor solutions are reported in the LinkOut documentation nor formally defined in any related data schema. We would also like to stress that since it is a CSV, a semi-structured data format, there is no explicit, complete and formal data schema. Therefore, to address those issues we had recourse to directly contacting the LinkOut service providers via e-mails, to exactly clarify the expected data exchange format by the LinkOut system. This is a time-consuming process, and while in this case it is notable that NCBI was very responsive, it is less

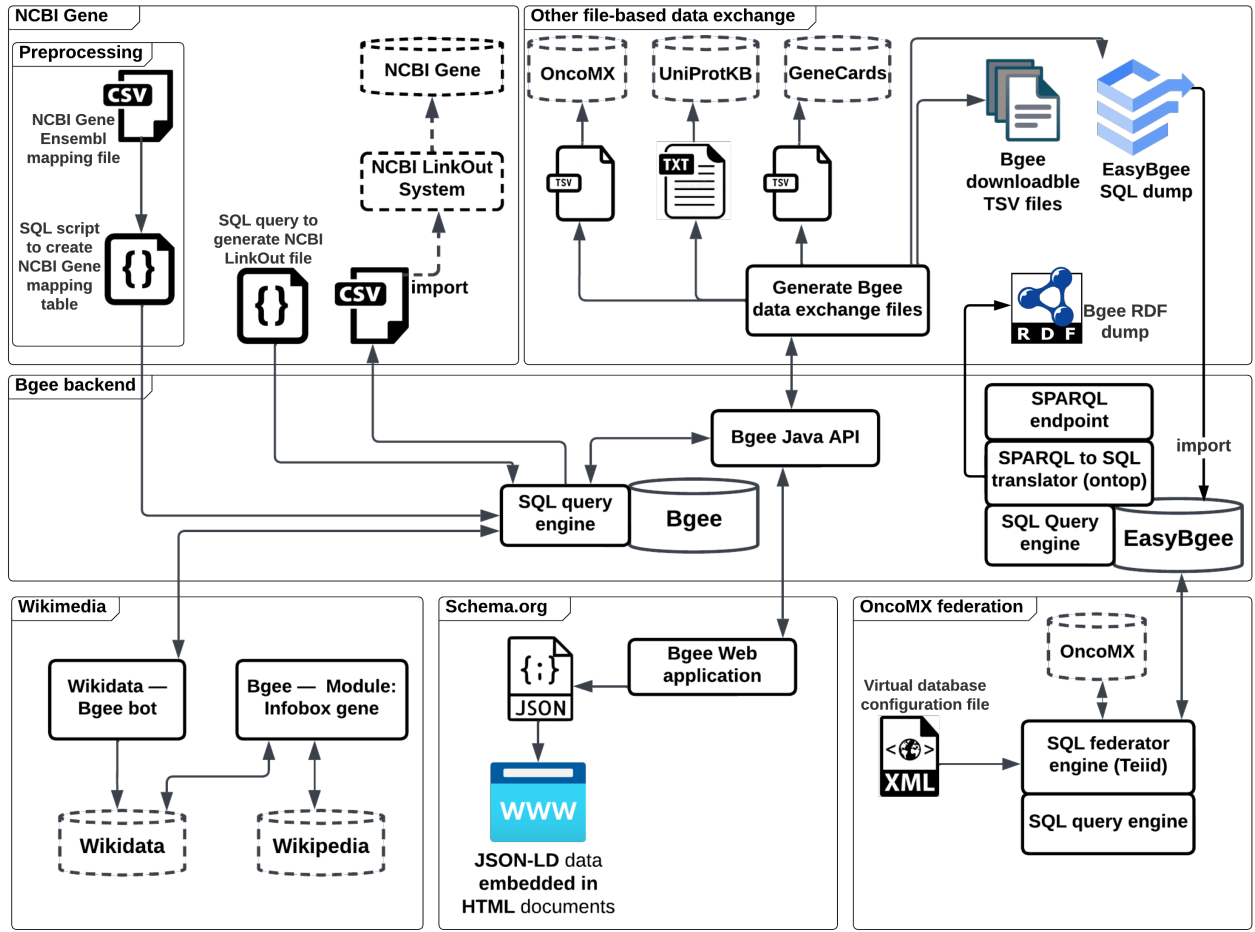

**Figure 2.** The architecture schema of the Bgee's interoperability with other major knowledge bases. Dashed cylinders represent external knowledge bases (KB). All rectangles are software components implemented by Bgee except the dashed one that is a third-party system component. In general, outgoing arrows from the "SQL query engine" and "Bgee Java API" are output data (e.g., retrieved results) and ingoing arrows to them are received SQL statements or API calls, respectively. All outgoing arrows from "Generate Bgee data exchange files" are file exports. The "NCBI Gene" container illustrates the components implemented and used to interoperate with NCBI Gene KB. "Other file-based data exchange" container groups all software components developed to generate Bgee exchange files specific to other KBs, non-KB specific TSV files containing views of the Bgee data, a MySQL database dump and a RDF dump of the Bgee data. To simplify the schema, not all KB interoperating with Bgee through files are shown, notably, RIKEN Metadatabase that imports the generated Bgee RDF dump; SPOKE, Open Systems Pharmacology and Monarch Initiative project that directly use the downloadable Bgee TSV files. "Bgee backend" container shows the data store layer and data access modes (i.e., Bgee Java APIs, SPARQL and SQL query languages) of the Bgee KB. "Wikimedia" container depicts the implemented software components to exchange information with Wikidata and Wikipedia. In the "Wikimedia" container, the arrow to Wikidata from the bot means data insertion from Bgee to Wikidata and the outgoing and ingoing arrows from/to "Module: Infobox Gene" represent querying and retrieving data, respectively. "Schema.org" container illustrates metadata are embedded in the Bgee Web pages with the Schema.org vocabulary, that are consumed by systems supporting this vocabulary such as Google Dataset Search. Finally, "OncoMX federation" container depicts a dynamic interoperability approach between two independent KBs (i.e., Bgee and OncoMX) that can replace file-based approaches (i.e., interoperability via static exchange files) as illustrated in the "Other file-based data exchange" container.

```
1  PrId,DB,UID,URL,IconUrl,UrlName,SubjectType,Attribute
2  10418,Gene,103476274,https://bgee.org/bgee15_0/gene/ENSPREG00000013759,,rbm41 gene expression,,
3  10418,Gene,103465305,https://bgee.org/bgee15_0/gene/ENSPREG00000013760,,kdm5ba gene expression,,
4  10418,Gene,103474996,https://bgee.org/bgee15_0/gene/ENSPREG00000013761,,BACH1 gene expression,,
```

**Figure 3.** A portion of the data exchange file used by the NCBI LinkOut system.

reliable than an available documentation.

Finally, with this use case we can see that although the exchanged information is simple (3), it is not straightforward to achieve interoperation between independent resources such as Bgee and NCBI Gene. The aforementioned issues would be significantly worse if the exchanged information were more complex, for example, including gene expression levels, anatomical structures and developmental stages, that are not expected by the LinkOut system. However, this relevant information would enrich, for instance, the existing "Expression" section of some NCBI Gene pages, and allow to further include this section for NCBI gene pages that do not have any expression information at time of writing, such as the chimpanzee's hemoglobin subunit beta gene page [55].

**Other KBs' experiences.** GeneCards and OMA developed a similar procedure to interoperate with the NCBI LinkOut system by adopting its one-side interoperability approach. In addition, UniProtKB is not exchanging information with the LinkOut system. Nevertheless, UniProtKB accession numbers are integrated and part of the NCBI Gene KB by applying a two-side interoperability approach as reported in [56]. For instance, the "human HBB" NCBI Gene page refers to UniProtKB accession numbers in Section "NCBI Reference Sequences (RefSeq)" and "Related Sequences" [57].

*Bgee in the UniProt knowledge base*

To interoperate with UniProt, we had to adopt its one-side interoperability approach that relies on a specific syntax and a text file format.

This file format can be interpreted as a CSV-like file where the separator is a semicolon followed with a space character (; ) and without a header row. Nevertheless, the first cell contains values defined with a specific syntax, that is a "[UniProt identifier] [internal code] [external resource name]" where white spaces are actually three times the space character, the character-set encoding is us-ascii and the internal code is composed of two letters. An example of a row (entry) in the UniProt information exchange format is shown below. In this example, the internal code used is "DR" that is the two letters code used by UniProt for cross-references.

```
A0A3B1E4W9   DR   Bgee; WBGene00304181; Expressed
    in pharyngeal muscle cell (C elegans) and 1
    other tissue.
```

Mainly thanks to the identifiers in the first and second columns, we are able to establish an interoperation between Bgee and UniProt. Consequently, for each UniProt protein entry, a corresponding Bgee gene entry is assigned with a link to a Bgee gene page. This link is built by prefixing the Bgee related identifier (e.g., `WBGene00304181`) with the Bgee gene page Web address `https://bgee.org/gene/`. Moreover, the third column contains the description we defined for each Bgee entry. This description is composed of the cell or tissue where the gene is highest expressed and of the number of tissues for which Bgee has expression data for this gene.

To generate the UniProt information exchange file, we developed a file writer that is part of the Bgee software and workflow. A new file is generated for each Bgee release. To assure that UniProt has access to the latest Bgee data, we provide a persistent URL [58].

**Other KBs' experiences.** OMA and GeneCards are also interoperating with UniProtKB by implementing the same approach as Bgee except that they do not use the description field to provide additional information as Bgee does.

*Bgee in the GeneCards knowledge base*
GeneCards is a KB that automatically integrates human gene-centric data from about 150 web sources, including genomic, transcriptomic, proteomic, genetic, clinical and functional information. Unlike NCBI Gene, GeneCards does not provide guidelines or a specific file format for information exchange. The absence of a predefined data format for interoperability gave us the freedom to define one and more flexibility about the information to exchange. First, before contacting GeneCards, we drafted a tab-separated values (TSV) file containing basic information from the Bgee database such as values that could be used as intersections between Bgee and GeneCards, Bgee gene page links, and short summaries about expression per gene. Moreover, to leverage our interoperability, we reuse existing data exchange workflows between Bgee and other KBs. For instance, the TSV generated for GeneCards contains similar information as the file exchanged with UniProtKB. The strategy we adopted was to first present a simple file with minimal information about Bgee entries that could be easily understood and included in the GeneCards' gene expression sections. With this strategy, our goal was to facilitate the discussions and to convince them to reuse our data. To further convince them, we also demonstrated our engagement and interest to publish links to their corresponding gene pages (or other data, if interest) in the Bgee website. Second, we contacted GeneCards (i.e., our potential data consumer) and presented them our solution for interoperability. Thanks to the simplicity, ease and benefits of adopting our proposed interoperability solution, GeneCards' maintainers quickly agreed with it, while suggesting a few changes. They then implemented a reader for this Bgee TSV file, resulting in the integration of gene expression information from Bgee as illustrated in Fig. 4. As a good practice, we also agreed to provide a persistent link pointing out to the TSV file containing the latest Bgee data [59]. Therefore, each new GeneCards release has access to the latest Bgee data.

Although we had to define how we would perform the information exchange between Bgee and GeneCards from scratch, to interoperate these two KBs did not require major efforts. For example, it required from us the presentation of a solution to an interoperability agreement, the exchange of five emails in total. The fact that the information being exchanged is minimal and simple was critical to the ease and rapidy of implementing an interoperation from scratch. Having established a prior data format exchange such as the NCBI LinkOut system enables to promote interoperability between KBs without requiring the implementation of a new data file reader each time a new resource appears to interoperate with. As a result, the interoperation is straightforward for the data consumer, once the data producer complies with the consumer's procedure and expectations. Therefore, the burden to perform interoperability is mostly put on the data producer side, i.e. one-side interoperability. The main drawback of the one-side interoperability is the lack of flexibility to add new information. For instance, with the LinkOut system we are not able to exchange a description such as a summary of gene expression, as it is the case with GeneCards where we established a two-side interoperability.

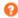

**Figure 4.** An example of a Bgee link and a gene expression summary in the GeneCards "Expression" tab.

**Other KBs' experiences.** UniProtKB information is present in GeneCards because GeneCards unilaterally extracts data from UniProtKB by using UniProtKB's one-side interoperability methods such as Web APIs [60]. On the other hand, OMA is not in GeneCards. However, OMA could establish a two-side interoperability approach similarly to Bgee and be part of the "Orthologs" and "Paralogs" sections for each GeneCards entry.

*Bgee in the OncoMX knowledge base*
The OncoMX is a KB that integrates relevant datasets to support the research of cancer biomarkers [40]. Bgee provides OncoMX with healthy gene expression data thanks to a two-side interoperability approach. As a result, the Bgee dataset is available in the OncoMX Web portal [61]. We, the Bgee team, have defined TSV data files that contain human and mouse gene expression present and absent calls per experiment condition (i.e., anatomical structures and developmental stages present in OncoMX) along with expression scores. Moreover, we reuse ontologies such as UBERON [62] for anatomical structures to avoid ambiguities and improve semantic interoperability. This facilitates the integration with other cancer biomarker related data such as the differential expression dataset that OncoMX integrates. In addition, a work towards a federated and automatic interoperability between Bgee and OncoMX has been done in the context of the Intelligent Open Data Exploration (INODE) project [63]. By applying this federated approach, we address most of the issues mentioned in the first paragraph of Subsection 4.1.

**Other KBs' experiences.** UniProtKB accession numbers (a.k.a. identifiers) are also assigned per biomarker or gene entry in OncoMX. This was done by the OncoMX developers when integrating different datasets and using mappings between gene names and UniProtKB identifiers. Therefore, although it is limited, UniProtKB and OncoMX are interoperating with a one-side interoperability approach. This could be improved by also querying UniProtKB to retrieve relevant information for OncoMX such as associated diseases to a given biomarker. Moreover, OncoMX could retrieve the GeneCards gene list and refer to the GeneCards entries (i.e., adding GeneCards cross-references) as it was done for UniProtKB. Alternatively, UniProtKB and GeneCards could build a ready to use dataset for OncoMX by applying a two-side interoperability method as Bgee did. Therefore, it would be more informative for the OncoMX users rather than a simple cross-referencing between KBs. Finally, OMA could provide human-mouse orthologs to relate the integrated human and mouse gene expression data from Bgee in OncoMX and directly give further insights for biomarker researchers in the OncoMX portal.

*Bgee database dumps and the RIKEN Metadatabase use case*
The Bgee data dumps that contain the main processed information are a simplified version of the entire Bgee relational database. As a result, we provide a simplified view that excludes the complexity of the integrated raw data by providing explicit and processed gene expression information. Without this view, it would be difficult for the end user (including third-party computer tools) to understand and deal with a massive amount of data and the writing of complex queries to extract the needed information. These data dumps contain highly structured data based on a relational data model and another one based on the Resource Description Framework (RDF) [64] data model. The relational database dump is called EasyBgee and Fig. 5 shows a portion of its data schema. We defined declarative mappings and applied the Ontop tool [65] to the EasyBgee database to generate the Bgee RDF dump [65, 66]. Therefore, the EasyBgee data are also available as RDF triple patterns, more specifically, using Turtle, the Terse RDF Triple Language, a concrete syntax for RDF [67].

We provide EasyBgee with both data models as a good practice to reach more users, to facilitate interoperability, and consequently, to make the Bgee data more reusable. For example, having the RDF dump available enabled the Japanese Institute of Physical and Chemical Research (RIKEN) to directly import the Bgee knowledge graph into the RIKEN Bioresource Metadatabase as a named graph. This RIKEN database integrates several life science datasets to support researchers in making a comprehensive use of RIKEN's research results. Thanks to this interoperation, RIKEN can now reuse Bgee gene expression data to support researchers when searching for a bioresource such as the use case for Alzheimer disease study described in [68].

**Other KBs' experiences.** Similarly to the Bgee use case, the RIKEN Metadatabase directly imports the available OMA RDF dump that is composed of fewer triples than the Bgee RDF dump. This cannot be the case of GeneCards, because it does not provide any RDF dump of its data. Furthermore, although importing the UniProtKB RDF dump to the RIKEN Metadatabase may be prohibitive because of its size (>100 billion of triples), UniProtKB provides views of its RDF data that could be reused by RIKEN, such as human diseases datasets.

*Bgee download files: a one-side interoperability approach*
As a one-side interoperability approach, Bgee provides views of its data as per-species TSV files. These files contain gene expression calls of presence/absence of expression and processed expression values that are currently used as the interoperability method with Bgee in different use cases. For example, the Monarch Initative project aims at connecting phenotypes (e.g., diseases) to genotypes (e.g., genes causing a disease) [42], and uses the Bgee download files to retrieve associations between genes, and the anatomical entities they are expressed in. This information is then displayed on their website in an "Anatomy" section of each gene page entry. Similarly, the Bgee download files are used to build a precision medicine open knowledge graph for a system called SPOKE [43], by retrieving association between genes and the anatomical entities where they are up- or down-regulated, to generate new edges in the graph. The download files are also used to build a knowledge base of gene expression information relevant to drug development in the context of the Open Systems Pharmacology suite [44], where Bgee provides a reference of normal gene expression in healthy conditions for multiple species, including human.

**Other KBs' experiences.** As a one-side interoperability approach, OMA, UniProtKB and GeneCards provide download files to exchange information including views of their data, respectively,

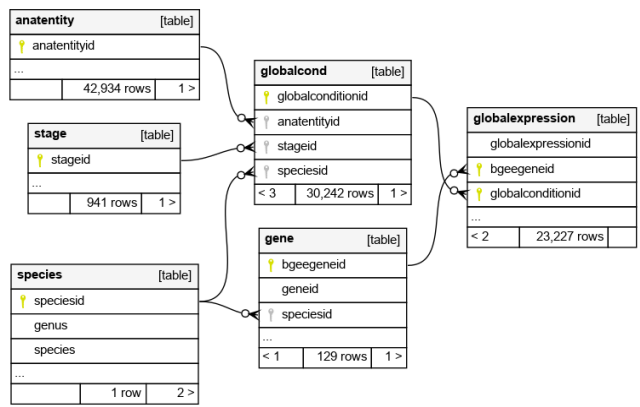

**Figure 5.** A portion of the EasyBgee relational data schema.

in [69], [70] and [71].

### 4.2 Programmatic interfaces

Providing several ways to programmatically interoperate and work with the data and information contained in a KB facilitates its reusability. This is because one interoperation method may be more suitable than another one depending on the user skills and use cases. In this regard, Bgee provides three distinct programmatic interfaces to query and to manipulate its data: a SPARQL endpoint, R packages, and a Web API. For the later one, although the Web API is already available [72], we are still working on providing a documentation and to be fully compliant with the OpenAPI [73] specification and standard to improve interoperability.

SPARQL is a structured query language and protocol for RDF-based data. Based on the Bgee RDF data dump depicted in subsection 4.1, a SPARQL endpoint is available in [74]. The query results can be retrieved in different formats such as JSON or CSV. Queries can be also executed though application programming interfaces (APIs) in several programming languages (e.g., Python via SPARQLWrapper [75]). Moreover, SPARQL also enables to perform federated join queries to combine various KBs that also provide SPARQL 1.1 endpoint such as Wikidata and UniProtKB. This capability of performing federated queries with Bgee is extensively demonstrated in [66].

The BgeeCall R package allows the user to generate present/absent gene expression calls without using an arbitrary cutoff (e.g. 1 TPM), by estimating background transcriptional noise based on non-expressed genomic features, i.e., intergenic sequences. We also provide the BgeeDB package for the annotation and gene expression data download from the Bgee database (i.e., interoperation), and for TopAnat analysis, a GO-like enrichment of anatomical terms, mapped to genes by expression patterns. Both packages along with their documentations are accessible at the Bgee website [76] and from Bioconductor [77]. Therefore, we reach Bioconductor's users that are interested in reusing gene expression related data. To facilitate the reuse of these packages, we make available a Docker container [78], a lightweight, standalone, executable package of software that includes all dependencies. In addition, other package repositories and systems interoperate with Bioconductor such as BioContainers [79]. As a result, we are able to attain a wider R user community. For example, so far, the BgeeDB package was downloaded around 325,000 times from BioContainers [80], that is significantly more than the 14,000 downloads from Bioconductor in the past 7 years.

**Other KBs' experiences.** UniProtKB and OMA provide SPARQL endpoints to access their data too. On the other hand, GeneCards does not have one. Having a SPARQL endpoint would facilitate, for example, the GeneCards interoperation with UniProtKB, OMA and Bgee as extensively demonstrated in [66]. In addition, OMA provides a R package in Bioconductor and REST APIs. Moreover, the OMA R package (OmaDB) was downloaded more than 67,000 times from BioContainers in the past 5 years. On the other hand, UniProtKB and GeneCards do not provide a R package authored and maintained by them, but UniProtKB makes available REST APIs. Both KBs could benefit bioinformaticians by providing a ready to use R package. We highlight that R language is one of the most used language in bioinformatics. To illustrate the relevance of doing this for reusability, a third-party R package that is not maintained by UniProtKB called "UniProt.ws: R Interface to UniProt Web Services" is available in Bioconductor [81] and has been downloaded more than 280,000 times from Biocontainers.

### 4.3 Automatising interoperability

The ideal interoperability of KBs is the one that allows seamless information exchange between them, in a way that looks like a unique system. To achieve this smooth and continuous information exchange, we aim for automatising interoperability. For example, we apply this approach to interoperate Bgee with Wikidata, Wikipedia, Orthologous MAtrix (OMA) [23], Google Dataset search engine [82], and OncoMX [83]. This approach addresses several issues of the file-based interoperability mentioned in Subsection 4.1, such as asynchronous and independent exporting and importing data operations.

To automatise interoperability between KBs, one of the steps in common between different solutions is to provide structured data, often by using interoperability standards to solve syntax and semantic heterogeneities. To do so, Bgee applies one-side, two-side, or multi-side interoperability depending on the use case, as described in the next paragraphs.

*Bgee in the Wikidata knowledge base*
Wikidata is an open and free KB that can be read and edited by any agent (i.e., both humans and machines). It contains and acts as a central storage of structured data related to other Wikimedia projects including Wikipedia. The Wikidata contents are available under a free license (i.e., CC0 [84]), can be exported using data standard formats, and can be interlinked to other data sets on the web of linked data. Its contents include highly relevant life science data. Thanks to the fact that Bgee data are also licensed under CC0, there is no restriction to reuse them in Wikidata. Fig. 6 shows a part of the Wikidata graph including Bgee data. To interoperate with Wikidata, we developed a bot that automatically extracts data from the Bgee relational database, structures them with the Wikidata data model [85], and loads them into the Wikidata KB. The bot is written in Python with the WikidataIntegrator library [86] and is available in our GitHub repository [45]. This bot inserts to Wikidata gene entries "expressed in" statements. For example, see the "expressed in" statements in the INS gene Wikidata page [87] and Fig.7 where one insertion is illustrated. Note that we defined versioned and persistent links as references to the "expressed in" statements, which is a good practice in order to track information provenance. Currently, only existing Wikidata gene entries from Ensembl and Wikidata anatomic entities (e.g., stomach) with a stated corresponding UBERON ontology term are considered (including Cell ontology). Thus not all data in Bgee is inserted into Wikidata. The Bgee gene entries for the species in common with Wikidata are identified with Ensembl gene ids. We do so to avoid ambiguities and to accurately include gene expression calls in Wikidata. Thus, the UBERON ontology and Ensembl gene identifiers allow us to address semantic heterogeneities.

Wikidata defines a specific data model to organise data and provides relations between properties in Wikidata and in RDF [85]. To interoperate with Wikidata, we must reuse existing Wikidata schemas or propose new ones based on the Wikidata model. Further instructions are available in [88]. Moreover, massive data insertion through a Wikidata bot such as the Bgee bot requires granted permissions by the Wikidata community [89]. Once this authorisation was granted, we were able to automatically insert and update the Bgee data in Wikidata entries. Permissions are often granted based on Wikidata contributors' support; currently, the Bgee bot is supported by three different Wikidata users [90]. Therefore, we perform a one-side interoperability because we had to strictly comply with the Wikidata procedure for interoperability. In addition, running the Bgee Wikidata bot is part of the Bgee pipeline final steps for each new release.

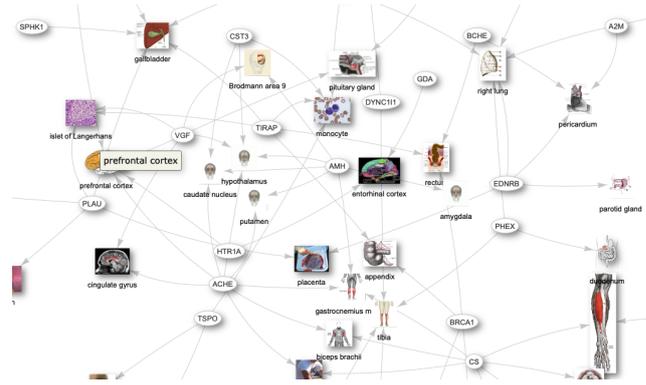

**Figure 6.** A portion of the Bgee data integrated into the Wikidata knowledge graph. It illustrates genes expression calls, where edges represent "expressed in" statements.

**Figure 7.** An "expressed in" statement entry in Wikidata by Bgee including provenance via a Bgee versioned URL. This image was extracted from the following Wikidata page https://www.wikidata.org/wiki/Q21163221.

**Other KBs' experiences.** GeneCards is a commercialised KB, UniProtKB and OMA are not CC0 licensed KBs, hence, they are not compatible with Wikidata's copyright requirements. Therefore, in principle, these KBs cannot interoperate with Wikidata. Nevertheless, non-CC0 KBs can donate part of their data under the CC0 license. Consequently, they can interoperate with Wikidata improving open knowledge reuse and increasing the traffic to these KBs. For example, currently, the presence of UniProtKB is limited to identifiers in Wikidata protein entries (i.e., as cross-references). These cross-references are fed to Wikidata with a third-party bot, *ProteinBoxBot* [91], thus, it is not maintained by UniProtKB.

*Bgee in Wikipedia*
The interoperation between Bgee and Wikipedia is fully automatic. From the end-user perspective, the anatomical structures such as pancreas where a gene is expressed along with links to the corresponding Bgee gene pages are included in the information box (infobox) of each Wikipedia gene article in English as illustrated in Fig. 8. To do so, we implemented a Lua [92] script that is defined in the Wikipedia gene infobox module [47], which retrieves structured data from Wikidata. Thus, this script queries Wikidata to fetch Bgee gene expression information and display it in the infobox. Since Bgee data were added by the Wikidata bot described above, the interoperation with Bgee is done indirectly through Wikidata. A highly relevant benefit of doing this is that changes in Wikidata are promptly available in the Wikipedia gene articles. Similarly to the Wikidata use case, changes in the Wikipedia infobox module code require permissions granted by the Wikipedia community and full compliance with their interoperability procedure, hence, an one-side interoperability approach. Nevertheless, a test environment so-called sandbox, for a given infobox module, is provided where, in principle, anyone can edit it [93].

**Figure 8.** A Wikipedia gene article contained gene expression information from Bgee. It was extracted from https://en.wikipedia.org/wiki/Insulin.

**Other KBs' experiences.** OMA is not present in Wikidata, hence, its data are not accessible by the Wikipedia Gene infobox module as in the Bgee use case. UniProtKB identifiers are referred in the Wikipedia gene pages' infobox thanks to their availability in Wikidata. GeneCards is also mentioned and links are built based on the gene names retrieved from Wikidata. However, none of these KBs (namely, OMA, GeneCards and UniProtKB) provide meaningful ready to use information for the Wikipedia users that is more than a simple external KB link. Moreover, the UniProtKB and GeneCards links in the Wikipedia Gene infoboxes are maintained by third-party contributors (i.e., non-authoritative source). Therefore, these KBs and other bioinformatics KBs can learn from this Bgee experience to improve their knowledge reuse.

*Bgee in the Google Dataset Search engine*

This use case is an example of a multi-side interoperability approach. Google Dataset Search engine fully automatises the process to index and to retrieve metadata from web pages that contain Schema.org structured data. Fig. 9 depicts a search of "homo sapiens gene expression" datasets in this Google tool. Notice that Schema.org is not exclusively under the authority of Google or Bgee. Therefore producing and consuming Schema.org structured data is in principle independent of the interoperable parts. This further allows other data consumers to reuse the data once they comply with the Schema.org approach. The compliance with a global data schema, data model (e.g., RDF graph), and syntax (e.g., JSON for Linked Data—JSON-LD [94]), intrinsically solves semantic and syntactical heterogeneities among interoperable parts. Founded by Google, Microsoft, Yahoo and Yandex, Schema.org vocabularies are developed by an open community process, using a mailing list [95] and through GitHub. Drawbacks of this approach include a lack of flexibility and that reaching an agreement for changes is not straightforward. Hence it greatly limits the information we are able to exchange. For instance, as of March 2nd, 2023, in the Schema.org GitHub, there are more than 700 issues open, some of them since 2014, and about 1300 closed [96].

**Figure 9.** Searching for human gene expression datasets and retrieving Bgee datasets via the Google Dataset search engine.

*Implementation details.* Google Dataset solely considers the Dataset, Datacatalog and Download concepts and their properties from Schema.org [97]. Therefore, interoperation with Bgee is restricted to these concepts. Although Taxon, Gene and "Anatomical structure" Schema.org concepts are not considered by Google Dataset Search, we also provide them via a JSON-LD embedded script at each Bgee gene page, and they can be consumed by any tool implementing this multi-side interoperability approach. Fig. 10 shows "expressed in" statements structured with Schema.org and included as a script in the human insulin Bgee gene page.

**Other KBs' experiences.** GeneCards and UniProKB do not use Schema.org to describe their datasets with metadata embedded in their Web pages. Therefore, datasets such as those in [70] are not directly available with Google Dataset Search. Similarly to Bgee, OMA implements Schema.org for datasets, but only a few of those listed in [69] are described. Moreover, OMA could provide richer metadata as Bgee has done, for instance, by assigning the downloadable forms that would allow users and software tools to directly access and download the files.

*Bgee and OncoMX: a knowledge base federation*

Federating data sources is the capacity of uniformly accessing data from distinct and potentially heterogeneous data sources without needing to physically move the data from them (i.e., data virtualization). Thus, the evaluation of queries are directly and real-time performed on the original data sources. By uniformly, we mean users see the data as if they were available in a single data source.

To address drawbacks of the file-based interoperability between Bgee and OncoMX as discussed in Subsection 4.1, we federate both KBs in the context of the INODE project [63], by applying a data virtualization and a one-side interoperability approaches. To implement this federation, we use Teiid [98], a real-time integration engine that supports a high query volume and transactions. The information to be exchanged with OncoMX is covered by EasyBgee, a materialised view of the Bgee relational database as described in

**Figure 10.** An example of the Bgee gene expression data structured with Schema.org.

Subsection 4.1 and its data schema illustrated Fig. 5. Thus in practice we interoperate OncoMX with EasyBgee. This is done in order to optimise query performance, because the native Bgee relational database does not explicitly provide the information needed in this federation. To define the virtual database (VDB) (i.e., the OncoMX and EasyBgee federation) with Teiid, we have to write an XML file that captures information about the VDB, the sources it integrates, and preferences for importing metadata [99]. The XML can also embed Data Definition Language (DDL) statements (i.e., some SQL commands). Fig. 11 depicts a portion of the XML file to set up a VDB based on OncoMX and EasyBgee. The OncoMX SQL database dump and the VDB XML file is available in [100].

As a result, the integration of those two data sources is no longer a manual effort, nor does it involve data duplication. Hence, the main advantage of federating by setting up a VDB is that whenever either Bgee or OncoMX gets updated, the new data is immediately available, as opposed to manual interoperation via TSV files. Moreover, it facilitates interoperability maintenance mainly because only relevant data modifications or changes at the data schema of data sources are required. It is often possible to perform needed changes by editing the VDB configuration, for instance, by creating SQL views as illustrated in Fig. 11. For example, currently, only human and mouse gene expression data are exchanged with OncoMX, nonetheless, if another species available in Bgee is of interest, an easy fix is to change the *SELECT* statement by adding the species taxon identifier in the listing (9606, 10090). Semantic and data heterogeneties may also be addressed by creating views and using SQL built-in functions to perform data transformations (e.g., concatenation of columns) during query evaluation. Therefore, keeping interoperation properly functional becomes an easier task, with no need to execute massive data export and import operations to deploy changes. In addition, we solve data storage heterogeneity as shown in Fig. 11, since EasyBgee is a native MySQL database and OncoMX a PostgreSQL one.

Finally, in the context of the INODE project [63], we provide an ontology-based data access to the OncoMX-Bgee federation with the Ontop tool. This is done mainly to achieve two aims: to improve the semantics of the data sources by applying ontologies to reduce ambiguities, and to provide the possibility of performing federated SPARQL queries, hence the capacity to interoperate with other knowledge graphs through its SPARQL endpoint [101].

**Figure 11.** A portion of the virtual database (VDB) configuration file set up to federate OncoMX and Bgee KBs. With the XML element `<model>`, we define the data source and metadata (e.g., data schema) to be considered. Some property/attribute values can also be assigned, for example, to set if a model should be visible or not in the federation. The metadata type asserted as `NATIVE` (i.e., `<metadata type="NATIVE"></metadata>` ) means the database metadata (e.g., data schema) will be considered exactly as it is originally defined in the data source. It is also with the `<metadata>` XML element we can create views, and consequently, new data schemas to structure the underlined data. This can be done through Data Definition Language (DDL) statements as shown above in the second `<metadata>` tag definition. The VDB XML file is fully documented in [99].

**Other KBs' experiences.** To the best of our knowledge, none of the KBs discussed in this article except from Bgee and OncoMX implements a federation over relational databases that are owned and managed by different organisations.

*Orthologous MAtrix (OMA) in Bgee*

Although we are focusing on the data provider side when considering two interoperable parts, in this subsection, we want to illustrate the interoperation from a data consumer perspective and to further justify the applicability of the interoperablity approaches discussed here by another KB too. OMA is a KB that contains information about evolutionary relationships among genes across species such as orthologs. Orthologs are genes in different species that evolved from a common ancestral gene by speciation. Bgee integrates OMA evolutionary relationships to further enable gene expression comparison among species.

To do so, Bgee interoperates with OMA through the OMA SPARQL endpoint. A tool along with several SPARQL queries based on the ORTHology ontology was developed to extract from the hierarchical orthologous group the pairwise orthology and paralogy relations [102]. The code source is available in the Bgee pipeline GitHub repository [103]. This tool is currently part of the Bgee pipeline for each new release assuring a semi-automatic procedure to exchange OMA information with Bgee. It is semi-automatic mainly because our one-side interoperability approach to include OMA data in Bgee is independent of OMA KB management and it does not allow real-time updates. Therefore, significant changes in OMA won't be fully automatically considered by our specialised tool. However, most of relevant changes can be done by editing the SPARQL queries or configuration-related files, for example, modifications in the OMA data schema (that does not happen often, last time was more than 2 years ago) may require to modify the SPARQL queries. Therefore, we facilitate interoperability maintenance and perform a quasi-seamless interoperability with OMA.

Finally, with this use case we want to emphasise that in an one-side interoperability approach either data producer or consumer may impose the way the interoperation is done. However, the fact of providing different ways to exchange information, bioinformatics KBs like OMA and Bgee, facilitates data reusability because it gives the possibility to choose which method is the more suitable by the data consumer. For example, OMA provides different programmatic interfaces such as Python libraries, REST APIs and SPARQL that further allows automatising interoperability. Therefore, the interoperability approach is imposed but with choices.

### 4.4 Final implementation considerations.

As a disclosure, when multiple options to interoperate with a target KB were available, we chose the one that was easier and faster for us to implement and did not compromise the minimal information we wanted to exchange. Otherwise, we implemented the only viable option for Bgee to be present in a target KB at that moment. For example, NCBI LinkOut system supports two data syntax, CSV and XML, we chose CSV because it was easier for us to execute a single SQL query over our relational database, and directly get the Bgee data in the expected tabular format for NCBI. Another example is related to our presence in Wikipedia that was first limited to including a few top organs where a human gene was expressed, that were already meaningful and valuable information for Wikipedia Gene pages. We did this to simplify discussion with Wikipedia moderators at first. Once a minimal interoperation was thus established between Bgee and Wikipedia, improvements were easier to perform and to deploy. Currently, Wikipedia also reuses Bgee mouse gene expression data, and cells are considered in addition to organs and tissues.

## 5 Ten lessons learned on improving the reusability of bioinformatics KBs

*Lesson 1—Partial interoperability is better than none.* We argue that it is better to have some extent of interoperability rather than none in order to increase (re)usability of a KB. A perfect and automatic interoperability between independent KBs is often hard to achieve due to issues such as legacy systems and practices, lack of resources (e.g., human resource allocation and skills, technology acquisition), technical incompatibilities, or reconciliation difficulties. Thus, aiming for partial interoperability when full interoperability is not feasible on the short term allows for delivering at least minimal data reuse along with the possibility to improve it over time.

*Lesson 2—Iteratively improving the information exchange is better than trying to achieve full interoperability at once.* Having less information to exchange can significantly simplify discussions between independent KB delegates and interoperability. A KB delegate will be more keen to perform an information exchange which is simple and easy to understand and implement than a complex one. This simplification usually results in partial interoperability but with the great benefit of being present and interoperating with a target KB. This target KB will be potentially more prone to accept and implement improvements later. Thus, once a minimal interoperation is established between KBs, improvements are easier to perform and to deploy.

*Lesson 3—Reusability implies better visibility, and vice-versa.* KB dissemination through information exchange with other KBs fosters better visibility, and consequently, more reusability. This is because data are reused not only by an external KB, but also, potentially, by its own user community and related software tools. Moreover, by providing provenance of the reused data, users may access and discover the original data source and enable more interoperability and data reuse.

*Lesson 4—Interoperability requires maintenance.* Similarly to a software development life cycle that includes maintenance, assuring long-term interoperability needs maintenance too. Therefore, interoperability efforts among KBs should ideally continue as long as they reuse data from each other. This also improves chances of reusing the latest data and of better quality and quantity of information exchanged. To reduce maintenance efforts, it is important to consider as a first step, before implementation, which interoperability approach is the most suitable given the requirements and constraints of the interoperable partners.

*Lesson 5—Automatise interoperability as much as possible.* There are various benefits of automatising interoperability, notably (a) reducing maintenance efforts, (b) providing real-time processing, and (c) facilitating scalability. An automatic approach significantly reduces maintenance efforts because once the data are up-to-date in a KB, the changes are propagated to other interoperable KBs and potentially at real time, for instance, via the execution of a bot. Moreover, real-time processing is an interoperability feature which the information received is processed by the data consumer almost immediately. To further explain (c), we can highlight KB federation approaches. By federating KBs, information exchange is easily scalable because adding a new KB or new type of information and data is done by editing configuration files and defining mappings between data sources (e.g., data schema alignments, data transformation functions).

*Lesson 6—Be flexible when choosing and providing interoperability approaches.* Defining how to establish interoperability depends on which methods are possible and available to the KBs. Moreover, different technical and resource constraints and KB delegates' skills may favour different approaches. For example, automatising interoperability requires KB delegate's technical skills that are not necessarily available. This can be alleviated by documentation (Lesson 10). In addition, a KB that makes one-side interoperability available should provide, if possible, distinct ways to exchange information. By doing so, a KB increases its reusability because it eventually matches users' skills and addresses third party KB constraints to achieve interoperation. Moreover, by having reusability as a main KB goal, the data producer or consumer should be prepared to make concessions and the target KB delegates are also less prone to collaborate, and consequently, to implement an interoperation, if they have to do more work. Thus, it is important to drastically reduce their work load when performing interoperability. For example, by providing ready-to-use data according to the target KB practices.

*Lesson 7—Focus on knowledge base delegates.* When a fully automated interoperability is not possible, a KB interoperation may only be possible, if it is preceded by communication between the delegates of KBs. This should lead them to collaborate with each other to interoperate their KBs. It is thus important to focus on establishing a good work relation between representatives of the interoperable KBs. Without this human reconciliation aspect, it can be difficult to get any technical interoperability mechanism to work. Moreover, any application of two-side interoperability approaches would not be possible without representatives' reconciliation.

*Lesson 8—There is a positive domino effect of knowledge base interoperability.* Interoperating with another KB can lead to a more and more complete network of information exchange among KBs, that is a positive "domino effect" for interoperability. This is thanks to three main reasons: (i) potential transitivity of interoperability, i.e., A interoperates with B that interoperates with C then A interoperates with C too; (ii) providing a positive example to convince further KB delegates to interoperate; and (iii) possible re-use of interoperability procedures. To further illustrate the later one, having already available interoperability solutions, such as a data virtualisation solution for federating KBs, leverage scalability for including new KBs in a KB interoperability network. Moreover, actively participating or collaborating with multi-side interoperability initiatives contributes to this positive domino effect. In summary, a continuous alignment with external resources propel data reusability.

*Lesson 9—Adopt the most appropriated license*. Although this lesson subject is already discussed in many distinct contexts including in the FAIR principles (i.e., "(Meta)data are released with a clear and accessible data usage license" [2]) we want to reinforce the importance of adopting as much as possible the least restrictive license for a KB. As a result, we eliminate possible legal interoperability issues, and hence, we can focus on addressing technical and human-aspect issues to exchange information. Nevertheless, we recognise the fact that some KBs need to apply highly restrictive licences. A reason for this might be to protect the authors' data ownership, but in fact hinders many possibilities for interoperability and open science [104].

*Lesson 10—Provide documentation, training, and tutorials for interoperability*. To leverage one-side interoperability approaches a KB should provide well-documented technical solutions, in-practice tutorials and training to facilitate reusability. Moreover, having communication channels with prompt responses is extremely important to assure a continuous user engagement, that includes external KB representatives.

## 6 Conclusion

To conclude, we argue that the best interoperability approach is the one that gets implemented despite partial interoperability or dissent between KB representatives. Therefore, one-, two- and multi-side interoperability methods are all relevant to promote KB data (re)use. Nevertheless, providing and implementing one or more of these methods should not be considered as final solutions for interoperability. This is because KBs, information exchange technologies and practices evolve, in addition to the apparition of new KBs. Finally, we illustrated with the Bgee KB several interoperability approaches and how we implemented them. We further illustrated how these approaches are transferable to other KBs by highlighting similar implementations by major bioinformatics KBs such as UniProtKB and what could be improved, when it is applicable. This allowed us to provide guidelines through pragmatic examples of how to interoperate with a variety of biological and general-purpose KBs such as Wikipedia.

## 7 List of abbreviations

API—Application Programming Interface
CSV—Comma-separated Values
DB—Database
DDL—Data Definition Language
DOI—Digital Object Identifier
DTD—Document Type Definition
FAIR—Findable, Accessible, Interoperable and Reusable
infobox—Information box
INODE—Intelligent Open Data Exploration
JSON—JavaScript Object Notation
JSON-LD—JavaScript Object Notation for Linked Data
KB—Knowledge Base
NCBI—National Center for Biotechnology Information
OMA—Orthologous Matrix
OWL—Web Ontology Language
RDF—Resource Description Framework
RIKEN—Japanese Institute of Physical and Chemical Research
SKOS—Simple Knowledge Organization System Reference
SPARQL—SPARQL Protocol and RDF Query Language
SWRL—Semantic Web Rule Language
SQL—Structured Query Language
TB—Terabytes
TPM—Transcripts Per Million
TSV—Tab-separated Values
Turtle—The Terse RDF Triple Language
UID—Unique Identifier
VDB—Virtual Database
VoIDext—Extended Vocabulary of Interlinked Datasets
XML—Extensible Markup Language
XSD—XML Schema Definition

## 8 Competing Interests

The author(s) declare that they have no competing interests.

## 9 Funding

SIB Swiss Institute of Bioinformatics; Canton de Vaud; Swiss National Science Foundation [173048, 207853, 167149]; NIH Award [U01CA215010 (OncoMX)]; European Union's Horizon 2020 research and innovation program [863410]; State Secretariat for Education, Research and Innovation (SERI) via ETHZ grant BG 02-072020. Funding for open access charge: Swiss National Science Foundation.

### 9.1 Author's Contributions

Conceptualization: TMF; Software: TMF, JW, FB; Investigation: TMF; Supervision: self-organisation; Writing original draft: TMF; Writing review and editing: all authors; Funding acquisition: TMF, FB, MRR.

## 10 Acknowledgements

We thank the representatives of all knowledge bases we are in contact with for continuously supporting the Bgee interoperability network. We also acknowledge the work of all past and present members of the Bgee team that made possible Bgee to be a successful knowledge base.